# Light: Particle & Wave

Eduardo Flores,[1]* Jeffrey Scaturro,[2]

[1]Physics & Astronomy Department, Rowan University, 201 Mullica Hill Rd., Glassboro, NJ 08028.
[2]Physics Department, Boston University, Boston, MA 00000.

*Correspondence to: flores@rowan.edu

**Keywords:** Double-slit experiment; Conservation laws; Particle-Wave duality

Light displays particle and wave properties within the bounds of the inequality, $K^2 + V^2 \leq 1$, where $K$ represents particle information and $V$ represents corresponding wave information. Using two laser beams that cross and then diverge from each other we explore the inequality. Results, based on semiclassical analysis, are in conflict with the inequality. However, analysis based on quantum electrodynamics resolves the conflict and uncovers a region where particle and wave aspects of light develop independently. Independent development of wave and particle aspects implies that they represent separate realities.

Common wave manifestation is associated with harmonic motion of many objects. An example is vibration of air molecules that results in sound. Particle manifestation is associated with properties of one small object. It is puzzling that electrons or photons could display particle and wave manifestations. Imagine a train of single photons aimed to a screen with two pinholes. Photons that go through the pinholes are collected on photosensitive film located away from the screen. The film shows random dots collected one at a time. A dot on the film represents particle aspect of the photon. The collection of dots forms a pattern similar to the pattern formed by a wave that goes through the pinholes. This behavior of light and matter is known as the wave-particle duality paradox. Interpretations of quantum mechanics [1] propose resolutions; however, there is no generally accepted one. According to Richard Feynman, wave-particle duality constitutes the only mystery of quantum mechanics [2]. This paper presents evidence that wave and particle aspects represent separate realities working together to display the observed behavior of light.

Wave and particle aspects of light can be quantified. The degree, to which particle trajectory is determined, is described by the path-information parameter ($K \leq 1$). Wave aspect is associated with formation of an interference pattern. Intensity contrast of the interference pattern is described by the visibility parameter ($V \leq 1$). Corresponding values of these parameters are limited by the Englert-Greenberger-Yasin inequality [3,4]

$$K^2 + V^2 \leq 1. \tag{1}$$

Most researchers consider this inequality as an expression of Bohr's principle of complementarity [5-7]. We note that there is no confirmed violation of this inequality.

Here, we present an experiment that challenges the inequality in Eq. (1). Taking advantage of a quantum property of coherent light we generate two independent beams from one source [8]. The setup is shown in Fig. 1. The source is a 50-mW Meredith Instruments laser that produces coherent light at wavelength of 632 nm.

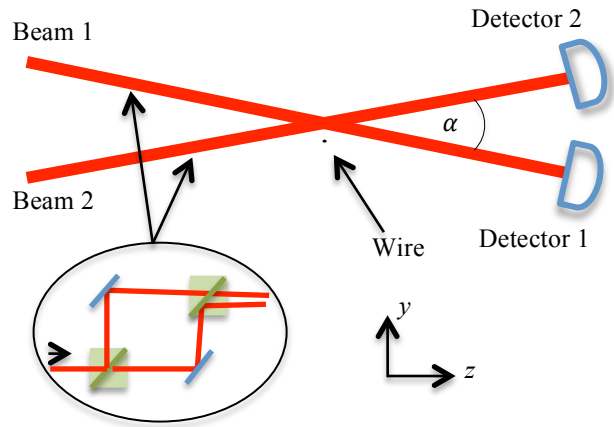

**Fig. 1. Experimental setup.** Using beam splitters and mirrors a laser beam generates two beams (inset). At the exiting beamsplitter, beam separation is 1.3 mm. The angle between the beams is $\alpha = 2.97$ mrad. The resulting beams cross each other and impinge upon separate detectors. A 17 $\mu$m thick wire, aligned parallel to the presumed interference fringes, is placed somewhere within the 1.0 mm beam intersection. The distance from beam splitter to wire is 0.454 m and from wire to detectors is 2.521 m.

When an opaque screen is placed at the beam intersection we see on it an interference pattern with high visibility ($V = 1$) but cannot determine path-information. On the other hand, when independent [8] beams cross freely, momentum conservation indicates that a photon that excites detector 1 (2) goes through the beam intersection along the path of beam 1 (2). Thus, path-information is maximal ($K = 1$) but we cannot determine visibility. To measure simultaneous visibility and path-information we place a 17$\mu$m thick wire somewhere across the beam intersection [9]. The effect of the wire on the beams is diffraction. In our analysis we use a semiclassical theory that consists of Fraunhofer diffraction, classical electrodynamics and elements of quantum physics. In the supplementary section we calculate the ratio $f = N/N_0$, where $N$ is photon count at end detector in the presence of the wire and $N_0$ is photon count without wire. Experimental data and theoretical calculation are in agreement as shown in Fig. 2 (**A**). Figure S1 shows that at low photon flux interference features in Fig. 2 (**A**) remain.

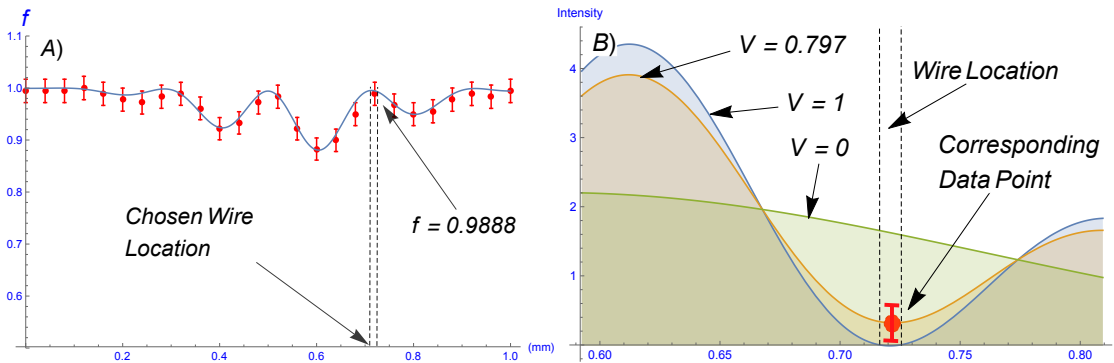

**Fig. 2. Experimental and theoretical results.** (**A**) Ratio $f = N/N_0$ is plotted as a function of wire position. Error bars are statistical. The solid line is the semiclassical prediction. Results from the other detector are essentially identical. The dashed lines show a data point with $f = 0.98883$ located at 0.72 mm. (**B**) We plot interference patterns in the neighborhood of chosen wire location (0.72 mm). The best fit corresponds to an interference pattern with visibility $V = 0.797$.

We use $f$ in Fig. 2 (**A**) to determine simultaneous path-information and visibility at the beam intersection. For instance, consider the data point at 0.72 mm. Since $f$ is 0.98883 then out of 100,000 photons 1,117 are lost. Analysis in the supplementary section shows that the wire blocks 588 photons and 588 photons are diffracted everywhere. Since only 59 out of the 98,883 photons that reach detectors are diffracted then Eq. (S15) gives high path-information, $K = 0.9994$. Using Eqs. (S9-S14) we obtain the visibility, $V = 0.797$. Therefore, we find a violation of the inequality in Eq. (*1*), as $K^2 + V^2 = 1.634 > 1$. We find similar results at other wire locations. We agree with researchers who argue [6,10-13] that quantum mechanical analysis of setups equivalent to the one in Fig. 1 does not leads to violation of the inequality in Eq. (1). We show that the violation of the inequality is due to limitations of the semiclassical model.

Considering the wire as a cylindrical potential barrier allows us to calculate wire diffraction in quantum electrodynamics (QED). The Feynman diagram for electron diffraction is shown in Fig. 3 (A). The calculation shows that transverse momentum transferred from wire to electron forms the diffraction pattern [14]. We note that when a direct measurement of wire radius (14 µm) is compared with estimates from classical diffraction (17 µm) we find that classical diffraction overestimates the wire radius [15]. However, the corresponding estimate from electron diffraction in QED gives the correct value. The photon case, shown in Fig. 3 (**B**), is similar to the electron case. However, the presence of the virtual electron loop makes this calculation challenging [14].

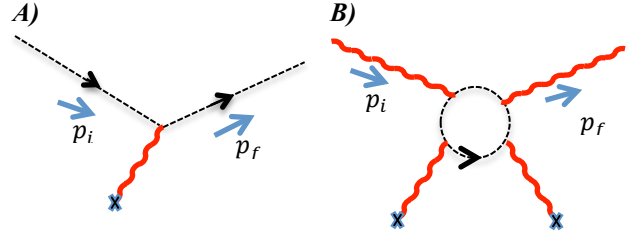

**Fig. 3. Feynman diagrams for a particle interacting with a potential.** (*A*) An electron interacts with static potential indicated by **x** at the end of the wavy photon line. (*B*) There are six diagrams for photon interacting with static potential but they are all related to this diagram.

The matrix element that corresponds to either Feynman diagram in Fig. 3 has the form [16]:

$$S_{if} = \delta_{if} - M_{if}, \qquad (2)$$

where $i$ and $f$ represent the initial and final state of the particle respectively. The delta function $\delta_{if}$ represents particles that go through unchanged. The amplitude $M_{if}$ describes particle-wire interaction. With the wire placed perpendicular to the incoming beam, scattering occurs on the plane perpendicular to the wire that contains the beam. On this plane the scattering angle is $\theta$. The amplitude depends on particle energy $(\kappa)$, $\theta$ and wire radius $(R)$.

$$M_{if}(\kappa, \theta, R). \qquad (3)$$

We analyze wire diffraction with interfering beams as in Fig. 1. We locate the wire at the center of the beam intersection ($y = 0$). The amplitude that a particle from either beam interacts with the wire is the sum of amplitudes [17],

$$S_{if} = \delta_{if} - \begin{cases} 0 & |y| > R \\ M_{if-} + e^{-i\Phi} M_{if+} & |y| \leq R \end{cases}, \qquad (4)$$

where $M_{if+}$ is the amplitude in Eq. (3) evaluated at $\left(\kappa, \theta + \frac{\alpha}{2}, R\right)$; this term corresponds to beam 1 in Fig. 1, and $M_{if-}$ evaluated at $\left(\kappa, \theta - \frac{\alpha}{2}, R\right)$ is the corresponding amplitude for beam 2. Equation (4) shows interference as a function of the phase difference ($\Phi$). Equation (4) also shows no interference outside the wire. Electron diffraction with interfering beams has been worked out in Ref. 14.

When the wire is removed, photon-photon scattering is the only interaction for the setup in Fig. 1; however, this cross-section is insignificant at the visible range [18]. Thus, Eq. (4) simplifies to $S_{if} = \delta_{if}$, which means that photons cross the beam intersection unperturbed. Momentum conservation requires that particle momentum distribution at the beam intersection is unchanged compared to the distribution before and after the beams cross. Therefore, particle momentum distribution at the beam intersection is simply

$$(p\hat{k}_1, p\hat{k}_2),$$ (5)

where $\hat{k}_1$ and $\hat{k}_2$ are unit vectors along beams 1 and 2 respectively and $p = \frac{h}{\lambda}$. On the other hand, classical time averaged field momentum density distribution at the beam intersection, given by

$$\langle \vec{g} \rangle = \epsilon_0 \langle \vec{E} \times \vec{B} \rangle = A_0 \cos^2\left[k(y - y_0)\sin\frac{\alpha}{2}\right]\hat{z},$$ (6)

displays maximum visibility ($V = 1$). The contrast between the quantum prediction in Eq. (5) and the classical prediction in Eq. (6) is graphically represented in Fig. 3 (**A**) in terms of energy-momentum distribution. Due to momentum conservation, path information is maximum ($K = 1$) for either theory. Thus, according to the semiclassical theory, with the wire removed, the violation of the inequality in Eq. (1) in reaches its maximum value ($K^2 + V^2 = 2$), while according to QED there is no violation.

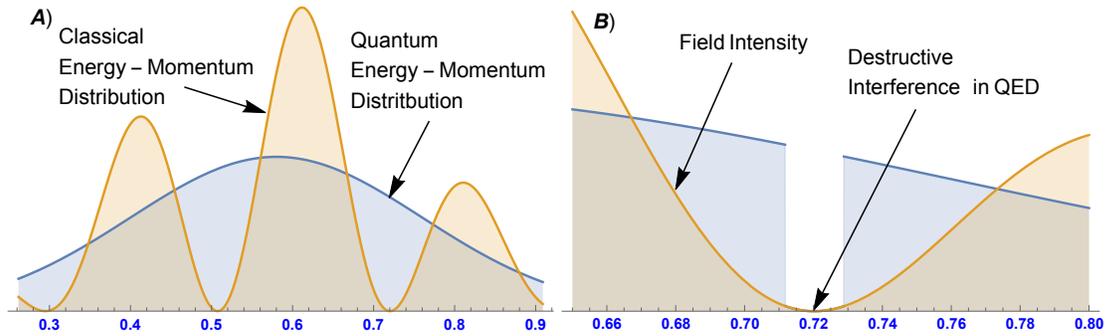

**Fig. 3. Classical and quantum predictions.** (**A**) Since the wire is not present, two beams with Gaussian profile cross freely. Quantum energy-momentum distribution shows no interference fringes ($V = 0$). Classical energy-momentum distribution shows constructive and destructive interference ($V = 1$). (**B**) The 17 $\mu$m thick wire is centered at 0.72 mm; now, at the wire there is destructive interference according to classical and quantum models.

Two fields in phase that cross at small angle form interference fringes. However, QED shows in Eq. (6) absence of momentum interference fringes for the experiment in Fig. 1 with the wire removed. In QED field effects are not fulfilled when the conservation laws cannot be simultaneously met. For instance, consider a high-energy photon propagating in vacuum; its field contains virtual electron-positron pairs. If the photon were to spontaneously decay into an electron and a positron there would be a violation of momentum conservation. To observe the virtual pair we need the nucleus of an atom to provide the necessary external momentum. Attempts to directly observe virtual pairs materialize them. In our experiment the wire plays a role similar to the nucleus of an atom when it provides the necessary external momentum to materialize the interference fringes in Fig. 2 (**A**). Attempts to directly observe interference fringes materialize them. Thus, it is based on linear momentum conservation that we believe that, for free photons, virtual electron-positron pairs and interference fringes are not physically present when not directly observed.

Presence of field interference fringes and lack of momentum interference fringes at the beam intersection for the experiment in Fig. 1 without the wire can have physical interpretation. Linear momentum, due to its conservation property, enables us to find photon trajectory. Trajectory is a particle property. Thus, linear momentum represents a particle aspect of the photon. On the other hand, the electromagnetic field represents a wave aspect of the photon. Therefore, at the beam intersection, particle aspect develops independently of wave aspect. Independent development of wave and particle aspects implies that wave aspect and particle aspect represent separate realities. Researchers working on the reality of the wave function could find this result useful [19-21]. In this paper we do not discuss the nature of wave aspect and particle aspect.

Classical electrodynamics predicts an interference pattern with high visibility across the beam intersection. This prediction seems to be confirmed experimentally as seen in Fig. 2 (**A**). Thus, photons that miss the wire and end at detectors would have high visibility ($V \approx 1$). In addition, due to momentum conservation and relatively small level of wire diffraction, path-information for photons that reach end detectors is high ($K \approx 1$) for semiclassical and quantum models. Therefore, according to the semiclassical theory there is a violation of the inequality in Eq. 1. However, according to Eq. (4), interference is localized at the wire as shown in Fig. 3 (**B**). Fortunately, this is enough to generate the observed pattern in Fig. 2 (**A**). We note that outside the wire there is no evidence of transverse momentum transfer to drive photons into an interference pattern. Since it is these photons, with high path information and zero visibility, which are measured by end detectors, we conclude that there is no violation of inequality in Eq. (1) in QED.

As an application we consider charged particles produced at high-energy particle colliders. After collision, these particles normally carry large amounts of energy-momentum. At first, these particles are free except for minor interactions with tracking devices. According to momentum conservation, free particles maintain their linear momentum *regardless* of the state of their wavefunction. Therefore, high-energy particles, at tracking devices, behave as classical particles. In fact, from tracks they leave on detectors, applying the conservation laws, we can infer with great accuracy particle properties: missing trajectory, decay time, charge, mass, etc. On the other hand, in atomic physics experiments, particles have more access to energy-momentum transfers; thus, wave effects are displayed since conservation laws are readily fulfilled.

**Supplementary Section:**

*Derivation of Ratio f*

We consider Babinet's principle applied to wire diffraction [22]. The electric field $\vec{E}_0$ of an undisturbed beam equals the electric field $\vec{E}_W$ produced by the same beam in the presence of the wire plus the electric field that would be produced by the beam in the presence of a hypothetical, complementary slit $\vec{E}_S$:

(S1)
$$\vec{E}_0 = \vec{E}_W + \vec{E}_S.$$

Outside the beam the original field is zero, $\vec{E}_0 = 0$, thus, according to Eq. (S1) we have that the field produced by the wire and the slit are similar except for sign, $\vec{E}_W = -\vec{E}_S$. We conclude that outside the beam, wire diffraction and complementary slit diffraction are identical.

Photon number conservation requires that

(S2)
$$N_0 = N_{PW} + N_{SW},$$

where $N_0$ is the total number of photons, $N_{PW}$ is the number of photons that go past the wire and $N_{SW}$ is the number of photons stopped by the wire. Since the wire and the slit are complementary constructs, the number of photons that go through the hypothetical slit, $N_{Slit}$, is the same as the number of photons that are stopped by the wire,

(S3)
$$N_{Slit} = N_{SW}.$$

From Eq. (*S1*) we obtain

(S4)
$$E_W^2 = E_0^2 - 2\vec{E}_0 \cdot \vec{E}_S + E_S^2.$$

The number of photons that go pass the wire is proportional to the integral of the field intensity in Eq. (S4) evaluated on a far away surface, large enough to catch all the photons,

$$N_{PW} = N_0 - 2 \int \vec{E}_0 \cdot \vec{E}_S \, da + N_{Slit}. \tag{S5}$$

Eq. (S2) together with Eq. (S3) and Eq. (S5) results in the identity

$$N_{SW} = \int \vec{E}_0 \cdot \vec{E}_S \, da. \tag{S6}$$

We note that $\vec{E}_0$, the field of the original beam, is non-zero in a small region entirely contained within the detector range. Integrating Eq. (S4) over region $D$ that covers just the original beam and using Eq. (S6) we have an expression for the number of photons in the beam

$$N = N_0 - 2N_{SW} + \int_D E_S^2 \, da. \tag{S7}$$

We note that the term $\int_D E_S^2 \, da$, where $\vec{E}_S$ is the field produced by the slit, represents the number of photons diffracted by the wire that fall within the beam. The field intensity ($E_S^2$) produced by a thin slit illuminated by a uniform field is the standard result [22]

$$E_S^2 \propto \mathrm{sinc}^2\left[\left(\frac{k\Delta y}{2}\right) \sin \theta\right], \tag{S8}$$

where $k = 2\pi/\lambda$, $\Delta y$ is the slit width and $\theta$ is the angle diffracted light makes relative to the direction normal to the slit.

Experimentally, a 5-mm diameter aperture placed in front of each detector defines region $D$. The transmitted light is focused onto the detector active region. We integrate the field intensity in Eq. (S8) over area $D$ to find the fraction of diffracted light that is collected by the detector:

$$\eta = 0.101. \tag{S9}$$

Thus, the number of photons that reach the detectors in Eq. (S7) is

$$N = N_0 - 2N_{SW} + \eta N_{Slit}. \tag{S10}$$

We define the ratio $f = N/N_0$. Using Eq. (S3) in Eq. (S10), we write

$$f = 1 - (2 - \eta)\frac{N_{SW}}{N_0}. \tag{S11}$$

The number of photons stopped by the wire ($N_{SW}$) is proportional to the flux that is blocked by the wire. The photons are linearly polarized along a common direction. According to classical electrodynamics, the field intensity at the beam intersection is a function of position ($y$), itself proportional to

$$e^{-\frac{1}{2}\left(\frac{y-y_0}{\sigma}\right)^2} \left\{a + \cos^2\left[k(y - y_0) \sin\frac{\alpha}{2}\right]\right\}, \tag{S12}$$

where the parameter $a$ allows for beams of different amplitude, $y_0$ is the offset of the pattern due to phase difference between beams, $\sigma$ is the Gaussian beam radius, $k = 2\pi/\lambda$, and $\alpha$ is the angle between the beams.

With the wire placed somewhere within the 1.0 mm beam intersection we integrate Eq. (S12) across the wire thickness ($\Delta y$) and obtain the number of photons stopped by the wire

(S13)
$$N_{SW}(y) = \Lambda \int_{y-\Delta y/2}^{y+\Delta y/2} e^{-\frac{1}{2}\left(\frac{y'-y_0}{\sigma}\right)^2} \left\{a + \cos^2\left[k(y'-y_0)\sin\frac{\alpha}{2}\right]\right\} dy',$$

where $\Lambda$ is a constant. We calculate $N_0$ by integrating Eq. (S13) across the entire beam intersection. The constant $\Lambda$, which also appears in $N_0$, is cancelled when calculating $f$ in Eq. (S11). Using Eq. (S13) and Eq. (S9) in Eq. (S11) we obtain $f$.

*Derivation of Visibility ($V$) and Path Information ($K$)*

We pick a data value for $f$ and using Eq. (S11) we obtain the ratio $N_{SW}/N_0$. According to Eq. (S13) $N_{SW}$ depends on $a$; thus, we find the value of $a$ that best matches $N_{SW}$. Once $a$ has been found we calculate the visibility using the formula

(S14)
$$V = \frac{1}{1+2a}.$$

We derive this formula by considering the interference of two plane waves of different amplitude that cross at a small angle. We obtain the intensity in Eq. (S12) with the exponential factor removed. Using the standard definition of visibility, $V = \frac{I_{max}-I_{min}}{I_{max}+I_{min}}$, we obtain Eq. (S14).

To calculate path information we consider the characteristics of photons that reach the detectors. Light that goes through the hypothetical slit complementary to the wire is diffracted light ($\vec{E}_S$). The slit thickness is the same as the wire thickness ($\Delta y$). In our setup, the wire thickness turns out to be $\Delta y = l/12.8$, where $l$ is the distance between two dark fringes. For two monochromatic light beams with a small angle $\alpha$ between them, $l$ is given by $l = \lambda/\alpha$, where $\lambda$ is the wavelength. A photon that goes through the slit is localized within the slit thickness, $\Delta y \leq \lambda/(12.8\alpha)$. As a result, the photon acquires transverse momentum with uncertainty $\Delta p_y \geq \hbar \frac{12.8\alpha}{2\lambda} = 6.4\hbar \frac{\alpha}{\lambda}$. On the other hand, the two original beams have momentum difference $(\vec{p}_2 - \vec{p}_1)_y = p\alpha$. Using the relation $p = \hbar k$, we see that the momentum difference between the photons from the original beams, $(\vec{p}_2 - \vec{p}_1)_y = \hbar \frac{2\pi\alpha}{\lambda} = 6.28\hbar \frac{\alpha}{\lambda}$, is of the same order as the uncertainty, $\Delta p_y \geq 6.4\hbar \frac{\alpha}{\lambda}$. Thus, the provenance of each photon diffracted by the slit is unknown and its path information is zero ($K = 0$). Therefore, photons diffracted by the slit have no path information. According to Babinet's principle in Eq. (S1) and the result in Eq. (S7) the slit and the wire produce similar diffracted fields; thus, we assert that photons diffracted by the wire also has no path information ($K = 0$).

We conclude that photons with full path information are the original number, $N_0$, less the number of photons stopped by the wire, $N_{SW}$, and all the diffracted photons, $N_{Slit}$, or, $N_0 - N_{SW} - N_{Slit}$. Using Eq. (S3) we write the number of photon with full path information as $N_0 - 2N_{SW}$. The ratio of photons with full path information to all the photons that reach the detector, $N$, in Eq. (S10) is our path information parameter,

(S15)
$$K = \frac{N_0 - 2N_{SW}}{N_0 - (2-\eta)N_{SW}},$$

where $\eta$ is given in Eq. (S9), $N_0$ and $N_{SW}$ are obtained from Eq. (S13). Eq. (S15) gives the correct values at the limit where the wire is large enough to stop half of the photons ($N_0/2$); in this case, according to Eq. (S3), the half that goes through are diffracted and have zero path information ($K = 0$), a result that agrees with Eq. (S15). If the wire is so small that does not stop photons, $N_{SW} = 0$, we expect the path information to be full ($K = 1$), which also agrees with Eq. (S15). We get an idea of the overall level of path information for our setup by averaging Eq. (S15) for different wire positions across the beam intersection and the result is $K_{Ave.} = 0.998$.

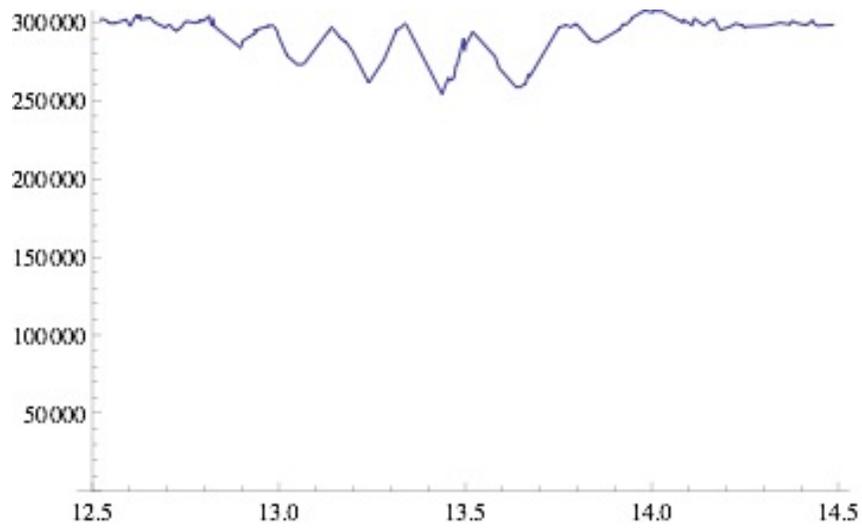

**Fig. S1. Low photon flux case.** The vertical axis shows photon count at end detector 1 as function of wire position. The position of the wire at the beam intersection is randomly chosen. There is strong evidence of an interference pattern with high visibility. The data collected by end detector 2 shows similar features. In this run, the average separation between one photon and the next is 3 km. Thus, it is likely that there is only one photon at a time in the entire setup.